\documentclass{article}
\usepackage{romp}
\usepackage{tensor}
\usepackage{mathtools}

\title{ Unified geometrical basis for the generalized Ehlers identities and Raychaudhuri equations }
\author{ Eduard G. Mychelkin$^1$ and Maxim A. Makukov$^2$ \\[2ex]
           \\ Fesenkov Astrophysical Institute, Almaty, Republic of Kazakhstan
           \\ $^1$e-mail: mychelkin@aphi.kz
            \\ $^2$e-mail: makukov@aphi.kz  }

\begin{document}

\maketitle
\begin{abstract}
It is shown that the timelike, spacelike and null versions of the Ehlers identity, as well as ensuing Raychaudhuri equations, might be all derived within a single geometrical approach based on the definition of the Riemann curvature tensor specified with respect to the corresponding congruence. Still, spacelike and null cases have a number of non-trivial peculiarities deserving special attention. 
\end{abstract}


\section{Introduction} 
\label{sec:introduction}

The  Ehlers identity and Raychaudhuri equation link kinematical characteristics of timelike congruences to the Ricci tensor and  energy-momentum tensor, correspondingly. They have a broad scope for applications starting from the seminal work of Raychaudhuri \cite{Raychaudhuri1955}. It is of special interest to consider possible analogues of these relations for other types of congruences -- spacelike and null. A deep geometrical insight into spacelike congruences with application to relativistic vortex hydrodynamics was gained by Greenberg \cite{Greenberg1970}. Few years ago, Abreu \& Visser \cite{Abreu2011} envisaged a number of phenomenological extensions of the timelike Raychaudhuri equation to spacelike and null congruences. Here we aim to develop \emph{ab initio} a single approach for the deduction of the Ehlers-Raychaudhuri relations of any type which proves to be appropriate for comparison and analysis of various arising geometrical and physical situations.

A terminological note is due. Quite often it is the expression 
\begin{equation}
{{R}_{\alpha \beta }}{{u}^{\alpha }}{{u}^{\beta }} = -\dot{\theta } - \frac{1}{3}{\theta }^{2}  -  2{{\sigma }^{2}} + 2{{\omega }^{2}} + {\dot{u}^{\alpha}}_{\,\,\, ;\alpha}.
\label{EhlersTrad}
\end{equation}
which is referred to as the Raychaudhuri equation (see, e.g., Hawking \& Ellis \cite{Hawking1973} and Wald \cite{Wald1984}). In a sense, this might be considered as an historical jargon, because in fact this expression is an explicit identity that follows algebraically from the Ricci identity (see below), and is satisfied by any metric (it becomes a proper equation only after the replacement of the Ricci tensor in accord with the Einstein equations).  Besides, in its universal covariant form commonly used today, the identity (\ref{EhlersTrad}) had been obtained by Ehlers \cite{Ehlers1961}, and that was acknowledged even by Raychaudhuri who referred to Ehlers' covariant result in \cite{De1968}. However, in non-covariant form specific to Friedmannian metrics the expression for $\tensor{R}{^0_0}$ was obtained earlier by Raychaudhuri \cite{Raychaudhuri1955}; another non-covariant form specific to stationary metrics was considered by Landau \& Lifshits \cite{Landau1994} (for additional historical notes, see, e.g., Kar \& Sengupta \cite{Kar2007}). In view of this, we will refer to the expression (\ref{EhlersTrad}) and its spacelike and null analogs as the Ehlers \emph{identities}. The result of the substitution of the Einstein equations for $\tensor{R}{_\alpha_\beta}$ in (\ref{EhlersTrad}) and its analogs will be referred to as the Raychaudhuri \emph{equations}.

\section{General algorithm}
\label{sec:gen}
Just as in the standard timelike case, we represent the generalized Ehlers identities as algebraic consequence of the Ricci identity 
\begin{equation}
{{v}_{\mu ;\nu \lambda }}-{{v}_{\mu ;\lambda \nu }}={{v}^{\alpha }}{{R}_{\alpha \mu \nu \lambda }},
\label{RiemannUSN}
\end{equation}
specified with respect to a given congruence $\{v^\mu\}$ which might be of time-like ($v^\mu v_\mu=1$), space-like ($v^\mu v_\mu=-1$) or null ($v^\mu v_\mu=0$) type, for space-time signature $\left(+--\;- \right)$ used in this paper. Videlicet, contracting over $\mu$ and $\lambda$ and projecting the result onto $v^\mu$-congruence, we get the {\textit{primary}} form of the generalized Ehlers identity:
\begin{equation}
{{R}_{\alpha \beta }}{{v}^{\alpha }}{{v}^{\beta }}={{v}^{\alpha }}_{;\beta \alpha }{{v}^{\beta }} - {v^\alpha}_{;\alpha\beta}v^\beta =    {\dot{v}^{\alpha}}_{\,\,\, ;\alpha}    -     {{v}_{\alpha ;\beta }}{{v}^{\beta ;\alpha }} -\dot{\theta} ,
\label{Ehlers}
\end{equation}
where $\theta = {{v}^{\alpha }}_{;\alpha }$ is expansion, overdot stands for Lagrangian derivative $ \dot{(\,\,)}= (\,\,)_{;\alpha} v^\alpha \equiv {{v}^{\alpha }}{{\nabla }_{\alpha }}$, and it is taken into account that 
${v^\alpha}_{;\beta\alpha}v^\beta \equiv  {\dot{v}^{\alpha}}_{\,\,\, ;\alpha}   -    {{v}_{\alpha ;\beta }}{{v}^{\beta ;\alpha }}.$  After substituting the Einstein equations,
\begin{equation}
{{R}_{\alpha \beta }}=8\pi G\left( {{T}_{\alpha \beta }}-\frac{1}{2}T{{g}_{\alpha \beta }} \right),
\label{EE}
\end{equation}
into (\ref{Ehlers}), we get the \textit{primary} form of the Raychaudhuri equation:
\begin{equation}
8\pi G({{T}_{\alpha \beta }}{{v}^{\alpha }}{{v}^{\beta }}-\frac{1}{2}T)={\dot{v}^{\alpha}}_{\,\,\, ;\alpha}    -     {{v}_{\alpha ;\beta }}{{v}^{\beta ;\alpha }} -\dot{\theta}.
\label{RaychBasic}
\end{equation}
Note that for the null case the term ${\dot{v}^{\alpha}}_{\,\,\, ;\alpha}$, as well as the trace ${{T}^{\alpha}}_{;\alpha}=T$ of the energy-momentum tensor, vanish (see below).

The next step is in revealing spacetime symmetries by splitting the term ${{v}_{\alpha ;\beta }}{{v}^{\beta ;\alpha }}$. For that we first use projections of covariant derivative operator $\nabla_\mu$ along the
$v^\mu$-congruence and onto the locally orthogonal 3-hypersurface by invoking the following generalized idempotent projection operator 
\begin{equation}
\tensor{p}{^\mu_\nu} = \tensor{\delta}{^\mu}_{\nu}-\epsilon v^{\mu}v_{\nu}, \qquad \tensor{p}{^\mu_\alpha} \tensor{p}{^\alpha_\nu} = \tensor{p}{^\mu_\nu},
\label{idemp}
\end{equation}
so that
\begin{equation}  
\nabla_\nu=\nabla_\nu^\parallel+\nabla_\nu^\perp=\epsilon v^\alpha v_\nu \nabla_\alpha+(\tensor{\delta}{^\alpha}_{\nu}-\epsilon v^{\alpha}v_{\nu})
\nabla_\alpha=\epsilon v^\alpha v_\nu \nabla_\alpha+\tensor{p}{^\alpha_\nu}
\nabla_\alpha \,,
\label{02proj}
\end{equation}
with the sign-factor $\epsilon \equiv v^{\alpha}v_{\alpha} = \{+1, -1, 0\}$ introduced to ensure the idempotence of $\tensor{p}{_\mu_\nu}$, including the degenerate null case.

Applying this to $v_\mu$-congruence, we get 
\begin{equation}  
v_{\mu ;\nu}=v_{\mu ;\nu}^\parallel+v_{\mu ;\nu}^\perp=\epsilon v_{\mu ;\alpha}v^{\alpha}v_{\nu}+v_{\mu ;\alpha}\tensor{p}{^\alpha_\nu}=\epsilon\dot{v}_{\mu}v_{\nu}+v_{\mu ;\alpha}\tensor{p}{^\alpha_\nu}\,.
\label{2proj}
\end{equation}
Now one splits the tensor $v_{\mu ;\nu}^\perp=v_{\mu ;\alpha}\tensor{p}{^\alpha_\nu}$ into symmetric and antisymmetric parts:
\begin{equation}
v_{\mu;\nu}^\perp = v^\perp_{(\mu;\nu)} + v^\perp_{[\mu;\nu]} \coloneqq \theta_{\mu\nu} + \omega_{\mu\nu}.
\label{decomp}
\end{equation}
Here the antisymmetric tensor\footnote{There are also other definitions (used for the timelike case) in the literature, $\tensor{\omega}{_\mu_\nu} = \tensor{p}{_\mu^\alpha}\tensor{p}{_\nu^\beta}\tensor{v}{_[_\alpha_;_\beta_]}$ and $\tensor{\omega}{_\mu_\nu} = \tensor{p}{_[_\mu^\alpha}\tensor{p}{_\nu_]^\beta}\tensor{v}{_\alpha_;_\beta}$, which are equivalent to (\ref{vort}).} is
\begin{equation}
\omega_{\mu\nu} = \frac{1}{2}\left( v^\perp_{\mu;\nu} - v^\perp_{\nu;\mu}  \right) =  \tensor{v}{_[_\mu_|_;_\alpha_|} \tensor{p}{^\alpha_\nu_]}, \quad \omega_{\alpha\beta}\omega^{\alpha\beta} \coloneqq 2\omega^2.
\label{vort}
\end{equation}
The symmetric tensor might be further split into trace-free and trace parts,
\begin{equation}
\theta_{\mu\nu} =  \frac{1}{2}\left( v^\perp_{\mu;\nu} + v^\perp_{\nu;\mu} \right) = \tensor{v}{_(_\mu_|_;_\alpha_|} \tensor{p}{^\alpha_\nu_)} = \sigma_{\mu\nu} + \theta {P}_{\mu\nu},
\label{exp}
\end{equation}
where $\theta$ is the trace of $\theta_{\mu\nu}$, and the normalized operator ${P}_{\mu\nu}$ should be proportional, according to (\ref{exp}),  to ${p}_{\mu\nu}$ and be of unit trace, i.e.: 
\begin{equation}
{P}_{\mu\nu} =  \frac{p_{\mu\nu}}{\tensor{p}{^\alpha_\alpha}}\,, \quad {{P}^\alpha}_{\alpha}=1.
\label{normOper}
\end{equation}
The tensor $\sigma_{\mu\nu}$ is the trace-free part of $\theta_{\mu\nu}$,
\begin{equation}
\sigma_{\mu\nu} = \theta_{\mu\nu} - \theta {P}_{\mu\nu}, \quad \sigma_{\alpha\beta}\sigma^{\alpha\beta} \coloneqq 2\sigma^2.
\label{shear}
\end{equation}
Note that introduction of the sign-factor  $\epsilon \equiv v^{\alpha}v_{\alpha} = \{+1, -1, 0\}$ in (\ref{idemp}) is justified taking trace of (\ref{2proj}) with simultaneous requirement for operator $p_{\mu \nu}$ to be idempotent. Then we obtain a simple constraint providing a unique solution:
\begin{equation}  
\tensor{v}{^\nu_{;\nu}}=\tensor{v}{^\nu_{;\alpha}}\tensor{p}{^\alpha_\nu}\, \Rightarrow \,\, \tensor{p}{^\alpha_\nu}= \tensor{\delta}{^\alpha}_{\nu}-\epsilon v^{\alpha}v_{\nu} \,,
\label{projector}
\end{equation}
where for the null case the projector proves to be trivial, $\tensor{p}{^\alpha_\nu}= \tensor{\delta}{^\alpha}_{\nu}$, because any null-vector is self-orthogonal.

Thus, an important role in the described algorithm is played by the projection operator $p_{\mu\nu}$ possessing the following properties:
\begin{enumerate}
\item symmetry: $p_{\mu \nu}=p_{\nu \mu}$,
\item idempotence: $p^\alpha_{\,\,\,\,\gamma} p^{\gamma}_{\,\,\,\,\beta}=p^\alpha_{\,\,\,\,\beta}$,
\item $\sigma$-orthogonality: $\sigma^\alpha_{\,\,\,\,\gamma} p^\gamma_{\,\,\,\,\beta}=0$.
\end{enumerate}
\noindent These properties prove to be valid for arbitrary congruences and space-time dimensions.

Taking into account (\ref{decomp}--\ref{exp}), the decomposition (\ref{2proj}) assumes the final form
\begin{equation}
v_{\mu ;\nu} = \omega_{\mu\nu} + \theta_{\mu\nu} + 	\epsilon\dot{v}_{\mu}v_{\nu} = \omega_{\mu\nu} + \sigma_{\mu\nu} + \theta {P}_{\mu\nu} + \epsilon\dot{v}_{\mu}v_{\nu}\,.
\label{2projFinal}
\end{equation}
All terms on the right side are mutually orthogonal, and the permutated square of the last term (i.e. $\dot{v}_{\mu}v_{\nu}\dot{v}^{\nu}v^{\mu}$) vanishes, so we immediately get: 
\begin{equation}
	{{v}_{\alpha ;\beta }}{{v}^{\beta ;\alpha }}=-2{{\omega }^{2}}+2{{\sigma }^{2}}+{{\theta }^{2}}\tensor{P}{_\alpha_\beta}\tensor{P}{^\beta^\alpha} = -2{{\omega }^{2}}+2{{\sigma }^{2}}+\frac{1}{\tensor{p}{^\alpha_\alpha}}{{\theta }^{2}}.
	\label{sq}
\end{equation}
Note that expression (\ref{sq}), unlike (\ref{2projFinal}), does not contain the sign-factor $\epsilon$ explicitly, but it sits latently in the operator $\tensor{p}{_\mu_\nu}$ contained in each term in (\ref{sq}).
Direct substitution of the term ${{v}_{\alpha ;\beta }}{{v}^{\beta ;\alpha }}$ into relations (\ref{Ehlers}) and (\ref{RaychBasic}) yields the resulting decomposed form of the generalized Ehlers identity and Raychaudhuri equation, correspondingly:
\begin{equation}
	{{R}_{\alpha \beta }}{{v}^{\alpha }}{{v}^{\beta }}={\dot{v}^{\alpha}}_{\,\,\, ;\alpha}    +2{{\omega }^{2}}-2{{\sigma }^{2}}-\frac{1}{\tensor{p}{^\alpha_\alpha}}{{\theta }^{2}} -\dot{\theta}
	\label{RprojGen}
\end{equation}
and
\begin{equation}
	8\pi G({{T}_{\alpha \beta }}{{v}^{\alpha }}{{v}^{\beta }}-\frac{1}{2}T)={\dot{v}^{\alpha}}_{\,\,\, ;\alpha}    +2{{\omega }^{2}}-2{{\sigma }^{2}}-\frac{1}{\tensor{p}{^\alpha_\alpha}}{{\theta }^{2}} -\dot{\theta}
	\label{RaychGen}.
\end{equation}

The basic algorithm (\ref{RiemannUSN}-\ref{RaychGen}) is general as it works on space-time of any dimension and signature, with ${p}_{\mu\nu}$ (and ${P}_{\mu\nu}$) to be specified separately for each type of congruence, as will be demonstrated below.

\section{Timelike case} 
\label{sec:timelike_case}

In timelike case $v^\mu=u^\mu$, $u^\alpha u_\alpha=1$, and from (\ref{idemp}) and (\ref{normOper}) it follows that $\tensor{{p}}{_\mu_\nu}={{g}_{\mu \nu }}-{{u}_{\mu }}{{u}_{\nu }} = {{h}_{\mu \nu }}$ (the standard notation for timelike projector), $\tensor{p}{^\alpha_\alpha}=3$ and ${P}_{\mu\nu} =  \frac{1}{3}h_{\mu\nu}$. Then, from (\ref{2projFinal}) we get the well-known decomposition:
\begin{equation}
{{u}_{\alpha ;\beta }}={{\omega }_{\alpha \beta }}+{{\sigma }_{\alpha \beta }}+\frac{1}{3}\theta {{h}_{\alpha \beta }}+{{\dot{u}}_{\alpha }}{{u}_{\beta }}\,.
\label{uabplus}
\end{equation}
For the opposite signature $\left(-++\;+ \right)$ instead of (\ref{uabplus}) one has ${{u}_{\alpha ;\beta }}={{\omega }_{\alpha \beta }}+{{\sigma }_{\alpha \beta }}+\frac{1}{3}\theta {{h}_{\alpha \beta }}-{{\dot{u}}_{\alpha }}{{u}_{\beta }}$, with ${{h}_{\mu \nu }}={{g}_{\mu \nu }}+{{u}_{\mu }}{{u}_{\nu }}$. In both signatures, explicit forms of vorticity tensor $\omega_{\mu \nu}$ (\ref{vort}) and shear tensor ${\sigma}_{\mu \nu}$ (\ref{shear}) (with corresponding ${{h}_{\mu \nu }}$) are:
\begin{equation}
{{\omega }_{\mu \nu }}\equiv \frac{1}{2}\left( {{u}_{\mu ;\alpha }}{{h}^{\alpha }}_{\nu }-{{u}_{\nu ;\alpha }}{{h}^{\alpha }}_{\mu } \right),
\label{wmunu}
\end{equation}
\begin{equation}
{{\sigma }_{\mu \nu }}\equiv \frac{1}{2}\left( {{u}_{\mu ;\alpha }}{{h}^{\alpha }}_{\nu }+{{u}_{\nu ;\alpha }}{{h}^{\alpha }}_{\mu } \right)-\frac{1}{3}\theta {{h}_{\mu \nu }}.
\label{smunu}
\end{equation}
With these, the relation (\ref{sq}) becomes:
\begin{equation}
{{u}_{\alpha ;\beta }}{{u}^{\beta ;\alpha }}=-2{{\omega }^{2}}+2{{\sigma }^{2}}+\frac{1}{3}{{\theta }^{2}},
\label{uab0}
\end{equation}
and in expressions (\ref{RprojGen}) and (\ref{RaychGen}) it should be adopted that $\tensor{p}{^\alpha_\alpha}=3$.
 
It is worth noting that (\ref{uab0}) should be distinguished from
\begin{equation}
{{u}_{\alpha ;\beta }}{{u}^{\alpha ;\beta }}=2{{\omega }^{2}}+2{{\sigma }^{2}}+\frac{1}{3}{{\theta }^{2}}+{{\dot{u}}_{\alpha }}{{\dot{u}}^{\alpha }}.
\label{uab1}
\end{equation}
Derivation of this expression is also straightforward taking into account the note after the formula (\ref{2projFinal}). The connection between ${{u}_{\alpha ;\beta }}{{u}^{\beta ;\alpha }}$ and ${{u}_{\alpha ;\beta }}{{u}^{\alpha ;\beta }}$ is then the following:
\begin{equation}
{{u}_{\alpha ;\beta }}{{u}^{\alpha ;\beta }}={{u}_{\alpha ;\beta }}{{u}^{\beta ;\alpha }}+4{{\omega }^{2}}+{{\dot{u}}_{\alpha }}{{\dot{u}}^{\alpha }},
\label{uab2}
\end{equation}
and so the primary Ehlers identity (\ref{Ehlers}) might also be represented in another useful form:
\begin{equation}
{{R}_{\alpha \beta }}{{u}^{\alpha }}{{u}^{\beta }}=  {\dot{u}^{\alpha}}_{\,\,\, ;\alpha}   -\dot{\theta} +4{{\omega }^{2}}+{{\dot{u}}_{\alpha }}{{\dot{u}}^{\alpha }}-{{u}_{\alpha ;\beta }}{{u}^{\alpha ;\beta }}.
\label{EhlersSymm}
\end{equation}

\section{Spacelike case} 
\label{sec:spacelike_case}

Unlike the timelike case which deals with bradyonic worldlines, spacelike congruences $\{v^\mu=n^\mu\}$, with $n_\alpha n^\alpha=-1$ and $n_\alpha u^\alpha=0$, might have various interpretations, such as tachyonic worldlines (e.g., as considered by Raychaudhuri himself \cite{Raychaudhuri1974}) or vorticity flows\footnote{The vortex tensor $\tensor{\omega}{_\mu_\nu}$ induces the dual vector flow $\omega^\mu = \sqrt{-g}\varepsilon^{\mu \alpha \beta \gamma} \omega_{\alpha \beta} u_\gamma$, which turns out to be spacelike, $\omega^\alpha \omega_\alpha = -1$.} in relativistic vortex hydrodynamics (see, e.g., \cite{Greenberg1970}). In such situations in accord with (\ref{idemp}) one finds that the idempotent operator $\tensor{p}{_\mu_\nu}=\tensor{\tilde{h}}{_\mu_\nu}$ of projecting onto 3-hypersurface orthogonal to $n^\mu$ takes the following form (we use tilde to denote spacelike case):
\begin{equation}
	\tensor{\tilde{h}}{_\mu_\nu}=\tensor{g}{_\mu_\nu}+n_\mu n_\nu, \quad \tensor{\tilde{h}}{^\mu_\alpha}\tensor{\tilde{h}}{^\alpha_\nu}=\tensor{\tilde{h}}{^\mu_\nu}, \quad \tensor{\tilde{h}}{^\alpha_\alpha}=3,
\label{SLproj}
\end{equation}
and, according to (\ref{normOper}), ${P}_{\mu\nu} =  \frac{1}{3} \tilde{h}_{\mu\nu}$. 
So, we obtain the primary and decomposed forms of the spacelike Ehlers identity:
\begin{equation}
	\tensor{R}{_\alpha_\beta}n^\alpha n^\beta = -\dot{\tilde{\theta}} + \tensor{\dot{n}}{^\alpha_{;\alpha}} -\tensor{n}{_\alpha_{;\beta}} \tensor{n}{^\beta^{;\alpha}}\,\,,
	\label{EhlersSL}
\end{equation}
\begin{equation}
{{R}_{\alpha \beta }}{{n}^{\alpha }}{{n}^{\beta }} = -\dot{\tilde{\theta} } - \frac{1}{3}{\tilde{\theta} }^{2}  -  2{{\tilde{\sigma} }^{2}} + 2{{\tilde{\omega} }^{2}} + {\dot{n}^{\alpha}}_{\,\,\, ;\alpha}\,\,,
\label{EhlersSLFin}
\end{equation}
with corresponding `shear' and `vorticity' tensors:
\begin{equation}
	{{\tilde{\sigma} }_{\mu \nu }}= \frac{1}{2}\left( {{n}_{\mu ;\alpha }}{\tilde{h}^{\alpha }}_{\nu }+{{n}_{\nu ;\alpha }}{\tilde{h}^{\alpha }}_{\mu } \right)-\frac{1}{3}\tilde{\theta}{\tilde{h}_{\mu \nu }},
	\label{sigmaSL}
\end{equation}
\begin{equation}
{{\tilde{\omega} }_{\mu \nu }}= \frac{1}{2}\left( {{n}_{\mu ;\alpha }}{{\tilde{h}}^{\alpha }}_{\nu }-{{n}_{\nu ;\alpha }}{{\tilde{h}}^{\alpha }}_{\mu } \right).
\label{vortSL}
\end{equation}

Despite formal similarity to timelike case, the interpretation here is entirely different because local 3-hypersurfaces orthogonal to spacelike congruence $\{n^\mu\}$ are of indefinite metric signature making it possible for the square of `shear' and `vorticity' tensors to have negative sign \cite{Abreu2011}. Besides, these hypersurfaces include, as seen from (\ref{SLproj}), time projections, and so, in general, are not unique for geometrical reasons.\footnote{For any spacelike vector $n_\mu$ locally orthogonal to timelike vector $u_\mu$, $n^\alpha u_\alpha = 0$, the vector $u^\mu$ is defined up to an additional vector $\lambda^\mu$, such that $u'^\mu = u^\mu + \lambda^\mu$ and $u_\alpha u^\alpha=u'_\alpha u'^\alpha= 1$, with $\lambda^\mu$ satisfying the conditions $\lambda_\alpha n^\alpha = 0$ and $\lambda_\alpha \lambda^\alpha + 2 \lambda_\alpha u^\alpha = 0$ \cite{Greenberg1970}.} Moreover, $\tensor{R}{_\alpha_\beta}n^\alpha n^\beta$ has geometrical meaning entirely different from $\tensor{R}{_\alpha_\beta}u^\alpha u^\beta$, so the corresponding decomposed spacelike Raychaudhuri equation,
\begin{equation}
8\pi G({{T}_{\alpha \beta }}{{n}^{\alpha }}{{n}^{\beta }}+\frac{1}{2}T)= -\dot{\tilde{\theta} } - \frac{1}{3}{\tilde{\theta} }^{2}  -  2{{\tilde{\sigma} }^{2}} + 2{{\tilde{\omega} }^{2}} + {\dot{n}^{\alpha}}_{\,\,\, ;\alpha},
\label{RaychSL}
\end{equation}
although mathematically correct, also loses customary  physical interpretation. 

On the other hand, if in solving a concrete problem one associates some spacelike congruence with a definite preferred spatial vector field $n^\mu$ (e.g., radial or axial), then the alternative (2+2)-decomposition might prove to be appropriate. In such cases one might employ the dyadic projection operator
\begin{equation}
\tensor{{p}}{_\mu_\nu}=\tensor{g}{_\mu_\nu}-u_\mu u_\nu + n_\mu n_\nu
\label{dyadicOper}
\end{equation}
with some preferred unit vectors $u^\mu$ and $n^\mu$. It is not difficult to show that (\ref{dyadicOper}) possesses the required properties of symmetry and idempotence, but its trace is $\tensor{{p}}{^\alpha_\alpha}=2$ rather than 1. Such situation has been considered in detail by Greenberg \cite{Greenberg1970} as applied to vortex hydrodynamics.

Next, the tetradic (1+1+1+1)-decomposition might also be appropriate for  spacelike congruences \cite{Greenberg1970}. For further development of Greenberg's methods see, e.g., \cite{Tsamparlis1983}.

\section{Null case} 
\label{sec:null_case}

For null congruence $\{v^\mu=k^\mu\}$ with $k_\mu k^\mu=0$ the idempotent operator according to (\ref{idemp}) and (\ref{projector}) becomes trivial (null congruence is locally orthogonal to itself):
\begin{equation}
{p^\mu}_{\nu}={\delta^\mu}_{\nu},
\label{nullop}
\end{equation}
and the normalized operator (\ref{normOper}) is found to be
\begin{equation}
{P}_{\mu\nu}=\frac{1}{4}p_{\mu\nu}=\frac{1}{4}g_{\mu\nu}.
\label{nullOper}
\end{equation}

Besides, as was noted earlier, in the null case we have ${\dot{k}^{\alpha}}_{\,\,\, ;\alpha} = 0$ because $\tensor{k}{^\alpha _;_\beta}k^\beta = {\dot{k}^{\alpha}} = 0$ by the definition of affine parameter for null geodesics, and so the general relation (\ref{Ehlers}) simplifies to the following primary form of the null Ehlers identity:
\begin{equation}
	{R}_{\alpha \beta }{{k}^{\alpha }}{{k}^{\beta }}= - {{k}_{\alpha ;\beta }}{{k}^{\beta ;\alpha }} -\dot{\theta},
\label{EhlersNull}
\end{equation}
where $\theta = {k^\alpha}_{;\alpha}$. Then from (\ref{sq}) and (\ref{RprojGen}) we obtain the decomposed form:
\begin{equation}
R_{\alpha\beta}k^\alpha k^\beta = -\dot{\theta} - 2\sigma^2 - \theta^2 {P}_{\alpha\beta} {P}^{\alpha\beta} + 2\omega^2 =  -\dot{\theta} - 2\sigma^2 - \frac{1}{4}\theta^2 + 2\omega^2,
\label{EhlersNullFinal}
\end{equation}
where, as usual, the shear term $-2 \sigma^2 $ includes the trace part $\theta^2{P}_{\alpha\beta} {P}^{\alpha\beta} = \frac{1}{4}\theta^2$. After substitution of the Einstein equations (\ref{EE}) with  zero trace, the identity (\ref{EhlersNullFinal}) transforms into the decomposed null Raychaudhuri equation: 
\begin{equation}
8\pi G{{T}_{\alpha \beta }}{{k}^{\alpha }}{{k}^{\beta }}= -\dot{\theta} - 2\sigma^2 - \theta^2 {P}_{\alpha\beta} {P}^{\alpha\beta} + 2\omega^2 =  -\dot{\theta} - 2\sigma^2 - \frac{1}{4}\theta^2 + 2\omega^2,
\label{RaychNullFinal}
\end{equation}
which might be applicable, for example, to the analysis of the Vaidya problem \cite{Vaidya1951,Mychelkin2017}.

The described approach to null fluids might be applied in spacetime of arbitrary dimension. There is no contradiction that (\ref{EhlersNullFinal}) appears to be different from the analogous expression used, e.g., by Hawking \& Ellis \cite{Hawking1973} and Wald \cite{Wald1984},
\begin{equation}
R_{mn}k^m k^n = -\dot{\hat{\theta}}  - 2\hat{\sigma}^2 - \frac{1}{2}\hat{\theta}^2 + 2\hat{\omega}^2,
\label{EhlersNullHE}
\end{equation}
because the latter was derived for effectively two-dimensional space of deviation vectors between null geodesics (with corresponding quantities denoted by hats). In its turn, applying our general algorithm to two-dimensional case we obtain the idempotent projector $\tensor{\hat{p}}{^m_n} = \tensor{\delta}{^m_n}$, and the corresponding normalized operator (\ref{normOper}) becomes
\begin{equation}
	\hat{P}_{mn}=\frac{\hat{p}_{mn}}{{\hat{p}^a}_{\,\,a}} = \frac{1}{2}\hat{p}_{mn}=\frac{1}{2}{\hat{g}}_{mn}, \quad \hat{P}_{m n}\hat{P}^{m n} = \frac{1}{2},
\end{equation}
with $m$, $n$ (and $a$) running over two values only. Then, according to (\ref{2projFinal}) with ${\dot{k}^{m}} = 0$, we obtain the decomposition
\begin{equation}
k_{m ;n}  = \hat{\omega}_{mn} + \hat{\sigma}_{mn} + \hat{\theta} {\hat{P}}_{mn} =  \hat{\omega}_{mn} + \hat{\sigma}_{mn} +\frac{1}{2} \hat{\theta}{\hat{g}}_{mn} \,,
\label{2projNull}
\end{equation}
where, in agreement with Hawking \& Ellis \cite{Hawking1973} and Wald \cite{Wald1984}, we have
\begin{equation}
\hat{\theta}={k}_{m;n} \hat{g}^{mn},
\end{equation}
\begin{equation}
\hat{\sigma}_{mn}= k_{(m; n)} - \hat{\theta} \hat{{P}}_{m n} = {k}_{(m;n)}-\frac{1}{2}\hat{\theta}\hat{g}_{mn}, \quad \hat{\sigma}_{m n}\hat{\sigma}^{m n} \coloneqq 2\hat{\sigma}^2,
\end{equation}
\begin{equation}
\hat{\omega}_{mn} = k_{[m; n]}, \quad \hat{\omega}_{m n}\hat{\omega}^{m n} \coloneqq 2\hat{\omega}^2.
\end{equation}
As a result, from (\ref{sq}) it follows
\begin{equation}
	{{k}_{m ;n }}{{k}^{n ;m }}=-2{\hat{\omega }^{2}}+2{\hat{\sigma }^{2}}+{\hat{\theta }^{2}}\tensor{P}{_m_n}\tensor{P}{^n^m} = -2{\hat{\omega }^{2}}+2{\hat{\sigma }^{2}}+\frac{1}{2}{\hat{\theta }^{2}},
	\label{sq0}
\end{equation}
and so in accord with (\ref{RprojGen}) we reproduce (\ref{EhlersNullHE}).

The Hawking-Ellis approach \cite{Hawking1973} was developed in the context of the singularity problem and  focusing theorem which correspond to representation of trace and trace-free parts of expansion in terms of null external curvature \cite{Bousso2016}. The latter implies existence of a null congruence emanating orthogonally to some spacelike 2-hypersurface. Our four-dimensional covariant algorithm (\ref{nullop})-(\ref{RaychNullFinal}) is also well defined. It leads to relations (\ref{EhlersNullFinal}) and (\ref{RaychNullFinal}) using only 4-dimensional quantities without, in general, singling out the quantities acting on 2-dimensional hypersurfaces.

\section{Conclusion} 
\label{sec:conclusion}

The Ehlers-Raychaudhuri relations deduced from the Ricci identity specified with respect to three types of congruence are represented in two distinct forms: primary and decomposed. The first is universal (functionally the same for any type of congruence), but further interpretation and applicability of the decomposed forms depends on the type of congruence under consideration.

For spacelike congruences the irreducible parts of (1+3)-decomposition lose their traditional physical interpretation. In this case, (2+2) and (1+1+1+1) splittings might be preferable, as shown, e.g., in the work on relativistic vortex hydrodynamics \cite{Greenberg1970} (see also in \cite{Raychaudhuri1974}).

As for null congruences, we have shown that our general algorithm might be used to obtain the 4-covariant as well as effectively two-dimensional representations.

It is of special interest to envisage non-normalized congruences $\{\xi^\mu \}$ with  $\xi_\alpha \xi^\alpha =\xi^2$. This allows to bring into consideration the symmetries of the so-called spacetime deformation tensor and to include into the Ehlers-Raychaudhuri relations a geometrical scalar $\xi$ which might be juxtaposed with certain physical scalar field \cite{Mychelkin2015}. We believe this opens a new scope for applications -- the issue we will consider in another publication.
 

\section*{Acknowledgement}
The work is performed within grant No. 0003-5/PCF-15-AKMIR.

\end{document}